\documentclass[twocolumn,preprintnumbers,amsmath,amssymb,pre,superscriptaddress,showpacs]{revtex4}
\usepackage{graphicx}
\usepackage{dcolumn}
\usepackage{bm}
\usepackage{psfrag}

\begin{document}
\renewcommand{\bbox}[1]{{\bm #1}}
\title{Staggered ladder spectra}
\author{E. Arvedson}
\affiliation{Department of Physics, G\"oteborg University, 41296
Gothenburg, Sweden}
\author{M. Wilkinson}
\affiliation{Faculty of Mathematics and Computing,
The Open University, Walton Hall, Milton Keynes, MK7 6AA, England}
\author{B. Mehlig}
\affiliation{Department of Physics, G\"oteborg University, 41296
Gothenburg, Sweden}
\author{K. Nakamura}
\affiliation{Department of Applied Physics, Osaka City University,
Sumiyoshi-ku, Osaka 558-8585, Japan}
\begin{abstract}
We exactly solve a Fokker-Planck equation by determining its
eigenvalues and eigenfunctions:  we construct nonlinear second-order
differential operators which act as raising and lowering
operators, generating ladder spectra for the odd and even
parity states. These are staggered: the odd-even separation
differs from even-odd. The Fokker-Planck equation describes, in
the limit of weak damping, a generalised Ornstein-Uhlenbeck
process where the random force depends upon position as well as
time. Our exact solution exhibits anomalous diffusion at short
times and a stationary non-Maxwellian momentum distribution.
\end{abstract}
\pacs{05.40.-a,05.45.-a,02.50.-r,05.40.Fb}
%
%
%
\maketitle
There are only few physically significant systems with ladder
spectra (exactly evenly spaced energy levels).
Examples are the harmonic oscillator and the Zeeman-splitting
Hamiltonian. In this letter we introduce a pair of exactly
solvable eigenvalue problems. We use them to solve two (closely
related) physically significant extensions of a classic problem in
the theory of diffusion, namely the Ornstein-Uhlenbeck process.
These eigenvalue problems have ladder spectra. Our systems differ
from the examples above in that their spectra consist of two
ladders which are staggered; the eigenvalues for eigenfunctions of
odd and even symmetry do not interleave with equal spacings. We
introduce a new type of raising and lowering operators in our
solution, which are nonlinear second-order differential operators.
Our generalised Ornstein-Uhlenbeck systems exhibit anomalous
diffusion at short times, and non-Maxwellian velocity
distributions at equilibrium; we obtain exact expressions which
are analogous to results obtained for the standard
Ornstein-Uhlenbeck process.

1. {\it Ornstein-Uhlenbeck processes}. Before we discuss our
extension of the Ornstein-Uhlenbeck process, we describe its usual
form. This considers a particle of momentum $p$ subjected to a
rapidly fluctuating random force $f(t)$ and subject to a drag
force $-\gamma p$, so that the equation of motion is $\dot
p=-\gamma p+f(t)$. The random force has statistics $\langle
f(t)\rangle=0$, $\langle f(t)f(t')\rangle=C(t-t')$ (angular
brackets denote ensemble averages throughout).  If the correlation
time $\tau$ of $f(t)$ is sufficiently short ($\gamma\tau \ll 1$),
the equation of motion may be approximated by a Langevin equation:
${\rm d}p=-\gamma p{\rm d}t+{\rm d}w$, where the Brownian
increment ${\rm d}w$ has statistics $\langle {\rm d}w\rangle=0$
and $\langle {\rm d}w^2\rangle=2D_0{\rm d}t$. The diffusion
constant is $D_0={1\over 2}\int_{-\infty}^\infty{\rm d}t \langle
f(t)f(0)\rangle$. This problem is discussed in many textbooks (for
example \cite{vKa92}); it is easily shown that the variance of the
momentum (with the particle starting at rest) is
\begin{equation}
\label{eq: 0} \langle p^2(t)\rangle=[1-\exp(-2\gamma t)]D_0/\gamma\,,
\end{equation}
that the equilibrium momentum distribution is Gaussian and that
the particle (of mass $m$) diffuses in space with diffusion
constant ${\cal D}_x=D_0/m^2\gamma^2$.

In many situations the force on the particle will be a function of
its position as well as of time. Our extension of the
Ornstein-Uhlenbeck process is concerned with what happens in this
situation when the damping is weak. We consider a force $f(x,t)$
which has mean value zero, and a correlation function $\langle
f(x,t)f(x',t')\rangle=C(x-x',t-t')$. The spatial and temporal
correlation scales are $\xi$ and $\tau$ respectively. If the
momentum of the particle is large compared to $p_0=m\xi/\tau$,
then the force experienced by the particle decorrelates more
rapidly than the force experienced by a stationary particle. Thus,
if the damping $\gamma $ is sufficiently weak that the particle is
accelerated to a momentum large compared to $p_0$, the diffusion
constant characterising fluctuations of momentum will be smaller
than $D_0$. The impulse of the force on a particle which is
initially at $x=0$ in the time from $t=0$ to $t=\Delta t$ is
\begin{equation}
\label{eq: 1}
\Delta w=\int_0^{\Delta t}{\rm d}t\
f(pt/m,t)+O(\Delta t^2) \ .
\end{equation}
If $\Delta t$ is large compared to $\tau$ but
small compared to $1/\gamma$, we can estimate $\langle \Delta
w^2\rangle=2D(p)\Delta t$, where
\begin{equation}
\label{eq: 2} D(p)=\frac{1}{2}\int_{-\infty}^\infty {\rm d}t\ C(pt/m,t)
\ .
\end{equation}
When $p\ll p_0$ we recover $D(p)=D_0$. When $p\gg p_0$,
we can approximate (\ref{eq: 2}) to obtain
\begin{equation}
\label{eq: 3} D(p)={D_1 p_0\over \vert p\vert}+O(p^{-2}) \ ,\
D_1={m\over 2p_0}\int_{-\infty}^\infty\!\!\!\! {\rm d}X\ C(X,0) \ .
\end{equation}
If the force is the gradient of a potential, $f(x,t)=\partial
V(x,t)/\partial x$, then $D_1=0$. In this case, expanding the
correlation function in (\ref{eq: 2}) in its second argument gives
$D(p)\sim D_3p_0^3/\vert p^3\vert$, where $D_3$ may be expressed
as an integral over the correlation function of $V(x,t)$. To
summarize: the momentum diffusion constant is a decreasing
function of momentum, such that $D(p)\sim  |p|^{-1}$ for a
generic random force, or $D(p)\sim |p|^{-3}$ for a
gradient force.

2. {\it Fokker-Planck equation}. The probability density for the
momentum, $P(p,t)$, satisfies a Fokker-Planck (generalised
diffusion) equation. Following arguments in \cite{vKa92}, one obtains
\begin{equation}
\label{eq: 4}
\partial_t P =\partial_p\big(\gamma p P
+D(p)\partial_pP\big) \ .
\end{equation}
Sturrock \cite{Stu66} introduced a related Fokker-Planck equation
(without the damping term) and also gave an expression for $D(p)$
analogous to equation (\ref{eq: 2}). His work was applied to the
stochastic acceleration of particles in plasmas, and subsequent
contributions have concentrated on refining models for the
calculation of $D(p)$ (see, for example, \cite{Ach91,Sta04}).
In the following we obtain exact solutions to
(\ref{eq: 4}) in the cases where $D(p)=D_1 p_0/|p|$ (which we
consider first) and $D(p)=D_3p_0^3/|p|^3$ (which can be treated in
the same way, and for which we discuss the results at the end of
the paper).

Introducing dimensionless variables ($t' = \gamma t$ and $z =  p
(\gamma/ p_0 D_1)^{1/3}$), the Fokker-Planck equation for the case
where $D(p)\propto 1/p$ becomes
\begin{equation}
\label{eq: 5} \partial_{t'} P =\partial_z\big(z P
+|z|^{-1}\partial_z P\big)\equiv \hat F P\,.
\end{equation}
It is convenient to transform the Fokker-Planck operator $\hat F$
to a Hermitian form, which we shall refer to as the Hamiltonian
operator
\begin{equation}
\label{eq: 6}
\hat H = P_0^{-1/2}\hat F P_0^{1/2} =
\frac{1}{2}-\frac{|z|^3}{4}+\frac{\partial}{\partial z}
\frac{1}{|z|} \frac{\partial}{\partial z}
\end{equation}
where $P_0(z)  \propto \exp(-|z|^3/3)$ is the stationary solution
satisfying $\hat F P_0 = 0$. We solve the diffusion problem by
constructing the eigenfunctions of the Hamiltonian operator. In
the following we make free use of the Dirac notation \cite{Dir30}
of quantum mechanics to write the equations in a compact form and
to emphasise their structure.

3. {\it Summary of principal results}. We start by listing our
results (for the case of random forcing, where $D(p)\propto 1/|p|$).
We construct the eigenvalues $\lambda_n$ and eigenfunctions
$\psi_n(z)$ of the operator $\hat H$. We identify raising and
lowering operators $\hat A^+$ and $\hat A$ which map one
eigenfunction to another with respectively one more or one fewer
nodes. We use these to show that the spectrum of $\hat H$ consists
of two superposed equally spaced spectra (ladder spectra) for even
and odd parity states:
\begin{equation}
\label{eq: 7} \lambda_n^+ = -3n\ , \ \ \lambda_n^- = -(3n+2)\ ,\ \
n=0\ldots \infty\ .
\end{equation}
The spectrum of the Hamiltonian (\ref{eq: 6}) is displayed on the
rhs of of Fig. 1. It unusual because the odd-even step is
different from the even-odd step, due to the singularity of the
Hamiltonian at $z=0$. Our raising and lowering operators allow us
to obtain matrix elements required for calculating expectation
values. One interesting result is the variance of the momentum for
a particle starting at rest at $t=0$:
\begin{equation}
\label{eq: 8} \langle p^2(t)\rangle =  \left(\frac{p_0
D_1}{\gamma}\right)^{2/3} \!\frac{3^{7/6}\Gamma(2/3)}{2\pi}
(1-{\rm e}^{-3\gamma t})^{2/3} \ .
\end{equation}
This is reminiscent of equation (\ref{eq: 0}) for the standard
Ornstein-Uhlenbeck process, however (\ref{eq: 8}) exhibits
anomalous diffusion for small times. At large times $\langle
p^2(t)\rangle$ converges to the expectation of $p^2$ with the
stationary (non-Maxwellian) momentum distribution
\begin{equation}
\label{eq: 9} P_0(p) = {\cal N}\exp\big(-\gamma |p|^3/(p_0
D_1)\big)
\end{equation}
(${\cal N}$ is a normalisation constant). At large times the
dynamics of the spatial displacement is diffusive $\langle
x^2(t)\rangle \sim 2 {\cal D}_xt$ with diffusion constant
\begin{equation}
\label{eq: 10} {\cal D}_x\!=\frac{(p_0D_1)^{2/3}}{m^2\gamma^{5/3}}
\frac{\pi 3^{-5/6}}{2\,\Gamma\!(2/3)^2} F_{32}
\Big(\frac{1}{3},\frac{1}{3},\frac{2}{3};\frac{5}{3},\frac{5}{3};1\Big)
\end{equation}
(here $F_{32}$ is a hypergeometric function). At small times, by
contrast, we obtain anomalous diffusion
\begin{equation}
\label{eq: 11} \langle x^2(t)\rangle = {\cal C}_x \,\bigl((p_0
D_1)^{2/3}\gamma^{-8/3}m^{-2}\bigr)\,{t}^{8/3}
\end{equation}
where the constant ${\cal C}_x$ is given by (\ref{eq: 31}) below.

\begin{figure}[t]
\includegraphics[width=3.75cm,clip]{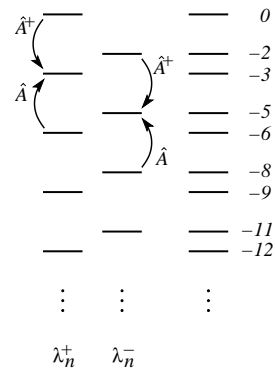}
\caption{\label{fig:1} The spectrum of $\hat H$ (right) is the sum
of two equally spaced (ladder) spectra $\lambda_n^-$ and
$\lambda_n^+$ shifted w.r.t each other (left). }
\end{figure}

4. {\it Ladder operators and eigenfunctions}. The eigenfunction of
the Fokker-Planck equation (\ref{eq: 4}) are alternately even and
odd functions, defined on the interval $(-\infty,\infty)$. The
operator $\hat H$, describing the limiting case of this
Fokker-Planck operator, is singular at $z=0$. We identify two
eigenfunctions of $\hat H$ by inspection, $\psi_0^+(z) = {\cal
C}_0^+\,\exp(-|z|^3/6)$ which has eigenvalue $\lambda_0^+ = 0$ and
$\psi_0^+(z) = {\cal C}_0^-\, z |z| \exp(-|z|^3/6)$ with
eigenvalue $\lambda^-_0 = -2$. These eigenfunctions are of even
and odd parity, respectively (zero and one node, respectively).
Our approach to determining the full spectrum will be to define a
raising operator $\hat A^+$ which maps any eigenfunction
$\psi_n^\pm(z)$ to its successor with the same parity,
$\psi_{n+1}^\pm(z)$, having two additional nodes.

We now list definitions of the operators we use: these include the
definitions of the raising and lowering operators, $\hat A^+$ and
$\hat A$, as well as an alternative representation of the Hamiltonian:
\begin{eqnarray}
\label{eq: 12} && \ \ \ \ \ \ \ \hat a^\pm = (\partial_z \pm
z|z|/2)
\nonumber \\
&&\hat A = \hat a^+ |z|^{-1} \hat a^+ \ ,\ \ \ \hat A^+ =
\hat a^- |z|^{-1} \hat a^- \nonumber \\
&&\hat H = \hat a^- |z|^{-1} \hat a^+ \ , \ \ \ \hat G = \hat a^+
|z|^{-1} \hat a^-\ .
\end{eqnarray}
Note that $\hat A^+$ is the Hermitian conjugate of $\hat A$. The
most significant properties of the operators $\hat A$ and $\hat
A^+$ are
\begin{equation}
\label{eq: 13} [\hat H, \hat A] = 3\hat A\quad \mbox{and}\quad
[\hat H, \hat A^+] = -3\hat A^+\
\end{equation}
(the square brackets are commutators). These expressions show that
the action of $\hat A$ and $\hat A^+$ on any eigenfunction is to
produce another eigenfunction with eigenvalue increased or
decreased by three, or else to produce a function which is
identically zero. The operator $\hat A^+$ adds two nodes,
and repeated action of $\hat A^+$ on $\psi_0^+(z)$ and $\psi_0^-(z)$
therefore exhausts the
set of eigenfunctions. Together with $\lambda_0^+ = 0$ and
$\lambda_0^- = -2$ this establishes that the spectrum of $\hat H$
is indeed (\ref{eq: 7}). Some other useful properties of the
operators of equation (\ref{eq: 12}) are
\begin{equation}
\label{eq: 14} [\hat A^+, \hat A] = 3(\hat H \!+\!\hat G) \ ,\ \ \hat
 H \!-\! \hat G=\hat I\ ,\ \ \hat A^+\hat A=\hat H^2\!+\!2\hat H \ .
\end{equation}
We represent the eigenfunctions by of $\hat H$ by kets
$|\psi_n^-\rangle$ and $|\psi_n^+\rangle$. The actions of $\hat A$
and $\hat A^+$ are
\begin{equation}
\label{eq: 15} \hat A^+ |\psi_{n}^\pm\rangle  =  C_{n+1}^\pm
|\psi_{n+1}^\pm \rangle \quad\mbox{and}\quad \hat A |\psi_{n}^\pm
\rangle =  C_{n}^\pm\!  |\psi_{n-1}^\pm \rangle \ .
\end{equation}
where (using (\ref{eq: 14})) we have $C_{n}^\pm =
\sqrt{3n(3n\mp2)}$.

\begin{figure}[t]
\mbox{}\\[5mm]
\includegraphics[width=7cm,clip]{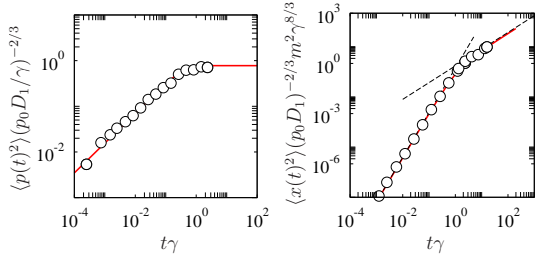}
\caption{\label{fig:2} Shows $\langle p^2(t)\rangle$ and $\langle
x^2(t)\rangle$. Computer simulation of the equations of motion
$\dot p = -\gamma p + f(x,t)$ and $m\dot x = p$ (symbols); theory,
eqs. (\ref{eq: 9}) and (\ref{eq: 26}), red lines. Also shown are
the limiting behaviours for $\langle x^2(t)\rangle$, (\ref{eq:
10}) and (\ref{eq: 11}), at long and short times (dashed lines).
In the simulations, $C(X,t) = \sigma^2
\exp[-X^2/(2\xi^2)-t^2/(2\tau^2)]$. The parameters were $m = 1$,
$\gamma = 10^{-3}$, $\xi = 0.1$, $\tau=0.1$, and $\sigma = 20$. }
\end{figure}

A peculiar feature of $\hat A$ and $\hat A^+$ is that they are of
second order in $\partial/\partial z$, whereas other examples of
raising and lowering operators are of first order in the
derivative. The difference is associated with the fact that the
spectrum is a staggered ladder: only states of the same parity
have equal spacing, so that the raising and lowering operators
must preserve the odd-even parity. This suggests replacing a first
order operator which increases the quantum number (total number of
nodes) by one with a second order operator which increases the
quantum number by two, preserving parity.

There is an alternative approach to generating the eigenfunctions
of (\ref{eq: 4}). This equation falls into one of the classes
considered in \cite{Inf51}, and we have written down first-order
operators which map one eigenfunction into another. However, these
operators are themselves functions of the quantum number $n$,
making the algebra cumbersome. We have not succeeded in
reproducing our results with the \lq Schr\" odinger factorisation'
method.

5. {\it Propagator and correlation functions}. The propagator of
the Fokker-Planck equation $\partial_{t'} K = \hat F K$ can be
expressed in terms of the eigenvalues $\lambda_n^\sigma$ and
eigenfunctions $\phi_n^\sigma(z) = P_0^{-1/2} \psi_n^\sigma(z)$ of
$\hat F$:
\begin{equation}
\label{eq: 16} K(y,z;t') = \sum_{n=0}^\infty \sum_{\sigma=\pm 1}
a_n^\sigma(y) \phi_n^\sigma(z) \exp(\lambda_n^\sigma t') \ .
\end{equation}
Here $y$ is the initial value and $z$ is the final value of the
coordinate. The expansion coefficients $a_n^\sigma(y)$ are
determined by the initial condition $K(y,z;0) = \delta(z-y)$,
namely $a_n^\sigma(y) = P_0^{-1/2}\psi_n^\sigma(y)$. In terms of
the eigenfunctions of $\hat H$ we have
\begin{equation}
\label{eq: 17} K(y,z;t') = \sum_{n\sigma} P_0^{-1/2}(y)
\psi_n^\sigma(y)
      P_0^{1/2}(z)\psi_n^\sigma(z)
\exp(\lambda_n^\sigma t')\,.
\end{equation}
The propagator determines correlation functions. Assuming $z_0 =
0$ we obtain for the expectation value of a function $O(z)$ at
time $t$
\begin{equation}
\label{eq: 18} \langle O(z(t'))\rangle
 = \sum_{n=0}^\infty
\frac{\psi_n^+(0)}{ \psi_0^+(0)} \langle \psi_{0}^+ |O(\hat
z)|\psi_{n}^+\rangle \exp(\lambda_n^+ t')\ .
\end{equation}
Similarly, for the correlation function of $O(z(t_2))$ and
$O(z(t_1))$ (with  $t_2'> t_1'> 0$)
\begin{eqnarray}
\label{eq: 19} \langle O(z(t_2')) O(z(t_1'))\rangle \!\!&=& \!\!
\sum_{nm\sigma} \frac{\psi_m^+(0)}{ \psi_0^+(0)} 
\langle \psi_{0}^+ |O(\hat z)|\psi_{n}^{\sigma}\rangle\nonumber\\
&&\hspace*{-2.5cm}\times
\langle \psi_{n}^\sigma |O(\hat z)|\psi_{m}^{+}\rangle 
\exp[\lambda_{n}^\sigma(t_2'-t_1')+\lambda_m^+ t_1']\,.
\end{eqnarray}
6. {\it Momentum diffusion}. To determine the time-dependence of
$\langle p^2(t)\rangle$ we need to evaluate the matrix elements
$Y_{0n} = \langle \psi_{0}^+ |\hat z^2|\psi_{n}^{+}\rangle$. A
recursion for these elements is obtained as follows. Let $Y_{0n+1}
= \langle \psi_{0}^+ |\hat z^2\hat
A^+|\psi_{n}^{+}\rangle/C^+_{n+1}$. Write $\hat z\hat A^+ = \hat z
\hat G + \hat z (\hat A^+-\hat G) = \hat z (\hat H-\hat I) +\hat z
(\hat A^+-\hat G)$. It follows
\begin{equation}
\label{eq: 21}
 \langle \psi_0^+|\hat z^2 \hat
A^+|\psi_n^+\rangle
 = (\lambda_n^+\!\!-1) Y_{0n}\! +\! \langle \psi_0^+| \hat z^2 (\hat A^+\!-\!\hat G)|\psi_n^+\rangle\,.
\end{equation}
Using $(\hat A^+-\hat G) = -\hat z\hat a^- $ and $[\hat z^3,\hat
a^-] = -3\hat z^2$ we obtain $Y_{0n+1}
=\lambda_n^++2Y_{0n}/C_{n+1}^+$, and together with $Y_{00} =
3^{7/6}\Gamma(2/3)/(2\pi)$ this gives
\begin{equation}
\label{eq: 22}
 Y_{0n} =  (-1)^{n+1} \frac{3^{17/12}
\Gamma(2/3)^{3/2}}{\sqrt{2}\pi^{3/2} (3n-2)
}\frac{\sqrt{\Gamma(n+1/3)}}{\sqrt{\Gamma(n+1)}}\,.
\end{equation}
We also find
\begin{equation}
\label{eq: 23} \psi_n^+(0)/\psi_{0}^+(0) =
(-1)^n\sqrt{\frac{\sqrt{3}\Gamma(2/3)}{2\pi}
\frac{\Gamma(n+1/3)}{\Gamma(n+1)}}
\end{equation}
and after performing the sum in (\ref{eq: 18}) we return to
dimensional variables. The final result is (\ref{eq: 8}).

7. {\it Spatial diffusion}. The time-dependence of $\langle
x_t^2\rangle$ is determined in a similar fashion, using (\ref{eq:
19}) and (\ref{eq: 23}):
\begin{equation}
\label{eq: 24} \langle x^2(t)\rangle =\frac{1}{\gamma^2}
\left(\frac{p_0 D_1}{\gamma}\right)^{2/3}\!\! \frac{1}{m^2}
\int_0^{t'}\!\!\!\!{\rm d}t_1'\int_0^{t'}\!\!\!\!{\rm d}t_2'\ \langle z_{t_1'}
z_{t_2'}\rangle\,.
\end{equation}
The matrix elements $Z_{mn} = \langle \psi_{m}^+ |\hat
z|\psi_{n}^{-}\rangle$ are found by a recursion method, analogous
to that yielding (\ref{eq: 22}):
\begin{eqnarray}
\label{eq: 25}
Z_{mn} &=& (-1)^{m-n} \frac{3^{5/6}}{6\pi}(m+n+1) \Gamma(2/3) \\
&\times&
\frac{\sqrt{\Gamma(n+1)\Gamma(m+1/3)}}{\sqrt{\Gamma(m+1)\Gamma(n+5/3)}}
\frac{\Gamma(n-m+1/3)}{\Gamma(n-m+2)}\,.
\nonumber
\end{eqnarray}
for $l\geq m-1$ and zero otherwise. Using (\ref{eq: 19})
\begin{equation}
\label{eq: 26} \langle x^2(t)\rangle =
\frac{(p_0D_1)^{2/3}}{m^2\gamma^{5/3}}
\sum_{k=0}^\infty\sum_{l=k-1}^\infty A_{kl} T_{kl}(t')
\end{equation}
with $A_{kl} = (\psi_k^+(0)/\psi_0^+(0)) Z_{0l} Z_{kl}$ and
\begin{eqnarray}
\label{eq: 27} T_{kl}(t')&=&  \int_0^{t'} \!{\rm d}t_1'
         \int_{t_1'}^{t'} \!{\rm d}t_2' \,
{\rm e}^{\lambda_l^-(t_2'-t_1')+\lambda_k^+t_1'}
\nonumber\\
&&+\int_0^{t'} \!{\rm d}t_1'
         \int_{0}^{t_1'} \!{\rm d}t_2'\,
         {\rm e}^{\lambda_l^-(t_1'-t_2')+\lambda_k^+t_2'}\,.
\end{eqnarray}
We remark upon an exact sum-rule for the $A_{kl}$, and also on
their asymptotic form for $k\gg 1$, $l\gg 1$:
\begin{equation}
\label{eq: 28} \sum_{k=0}^l A_{kl}=0 \ ,\,\  A_{kl} \!\sim\!
\frac{\Gamma(2/3)^2}{3^{1/3} 4\pi^2}\frac{k+l}{k^{2/3} l^{4/3}
(l-k)^{5/3}}\,.
\end{equation}
We now show how to derive the limiting behaviours (\ref{eq: 10})
and (\ref{eq: 11}), shown as dashed lines in Fig. \ref{fig:2}. At
large time $x$ evolves diffusively: $\langle x^2\rangle \sim
2{\cal D}_xt$, with the diffusion constant obtained from the
equilibrium momentum correlation function:
\begin{eqnarray}
\label{eq: 29} {\cal
D}_x&=&{1\over{2m^2\gamma}}\biggl({p_0D_1\over{\gamma}}\biggr)^{2/3}
\lim_{T\to \infty}\int_{-\infty}^\infty {\rm d}t'\ \langle
z_Tz_{t'+T}\rangle \nonumber \\
&=&{-1\over{2m^2\gamma}}\biggl({p_0D_1\over{\gamma}}\biggr)^{2/3}
\sum_{n=0}^\infty {Z_{0n}^2\over{\lambda^-_n}}
\end{eqnarray}
which evaluates to (\ref{eq: 10}). 
At small values of $t'$ the double sum 
(\ref{eq: 26}) is dominated by the large-$k,l$ terms. We thus evaluate the
small-$t'$ behaviour by approximating the sums in (\ref{eq: 26})
by integrals  and using the asymptotic form for the coefficients
$A_{kl}$. There is a non-integrable divergence of $A(k,l)$ as
$k\to l$, which can be cancelled by using the sum rule in equation
(\ref{eq: 28}). We obtain the limiting behaviour (\ref{eq: 11})
with
\begin{eqnarray}
\label{eq: 31} {\cal C}_x&=& -C\,
\int_0^\infty\!\!\! {\rm d}x\ x^{-8/3}\\
\nonumber &\times&\int_0^1\!\! {\rm d}y\
\biggl[{a(x)-a(xy)\over{1-y}}+xa'(x)\biggr]
{1+y\over{y^{2/3}(1-y)^{5/3}}}
\end{eqnarray}
\mbox{}\\[-1mm]
where $a(x)=[1\!-\!\exp(-x)]/x$ and $C = 3^{-7/3} \Gamma(2/3)/(2
\pi^2)$. The integral is convergent and can be evaluated
numerically to give ${\cal C}_x =0.57\ldots $. This is in good
agreement with a numerical evaluation of the sum (\ref{eq: 26}),
as shown in Fig. \ref{fig:2}.

8. {\it Gradient-force case}. In the case where the force is the
gradient of a potential function, we have $D(p) = D_3
p_0^3/|p|^3+O(p^{-4})$ \cite{Gol91}. In dimensionless variables
the Fokker-Planck equation is
\begin{equation}
\label{eq: 32} \partial_{t'}P={\partial_z}\big(zP
+|z|^{-3}\partial_z P\big)\equiv \hat F P
\end{equation}
instead of (\ref{eq: 5}). This Fokker-Planck equation has the
non-Maxwellian equilibrium distribution $P_0(z) = \exp(-|z|^5/5)$.
The raising and lowering operators are of the form $\hat A^+=\hat
a^- |z|^{-3}\hat a^-$ and $\hat A = \hat a^+ |z|^{-3} \hat a^+$
with $\hat a^\pm = (\partial_z\pm z|z|^3/2)$. The analogue of
(\ref{eq: 13}) is $[\hat H,\hat A]=5\hat A$, $[\hat H,\hat
A^+]=-5\hat A^+$, and the eigenvalues are
$0,-4,-5,-9,-10,-14,-15,\ldots$. In this case, too, a closed
expression for example for $\langle p^2(t)\rangle$ can be obtained
which is analogous to (\ref{eq: 0}), but which exhibits anomalous
diffusion:
\begin{eqnarray}
\label{eq: 33} \langle p_{t}^2\rangle &=& \left(\frac{p_0
D_3}{\gamma}\right)^{2/5}
\frac{5^{2/5} \sin(\pi/5)}{\pi}\\
&\times&\Gamma(3/5)\Gamma(4/5)\big(1-{\rm e}^{-5\gamma t})^{2/5}\,.
\nonumber
\end{eqnarray}
The short-time anomalous diffusion is consistent with the
asymptotic scaling obtained in \cite{Gol91,Ros92} for the
special case of undamped stochastic acceleration.

A full account of our results, and their extension to higher
dimensions, will be published elsewhere.

9. {\it Acknowledgements}. We thank Stellan \"Ostlund for
illuminating discussions. MW thanks JSPS for a visiting
fellowship. 

\vfill\flushbottom
\end{document}